# Detecting Band Profiles of Devices with Conductive Atomic Force Microscopy

*Ranran Li,[1,3,4] Takashi Taniguchi,[2] Kenji Watanabe,[2] Jiamin Xue[1]\**

[1]School of Physical Science and Technology, ShanghaiTech University, Shanghai 201210, China.

[2]National Institute for Materials Science, Tsukuba 305-0044, Japan.

[3]University of Chinese Academy of Sciences, Beijing 100049, China

[4]Shanghai Institute of Ceramics, Chinese Academy of Sciences, Shanghai 200050, China

\*Author to whom correspondence should be addressed: xuejm@shanghaitech.edu.cn

**ABSTRACT**

Band profiles of electronic devices are of fundamental importance in determining their properties. A technique that can map the band profile of both the interior and edges of a device at the nanometer scale is highly demanded. Conventional scanning tunneling spectroscopy (STS) can map band structure at the atomic scale, but is limited to the interior of large and conductive samples. Here we develop a contact-mode STS based on conductive atomic force microscope that can remove these constraints. With this technique, we map the band profile of $MoS_2$ transistors with nanometer resolution at room temperature. A band bending of 0.6 eV within 18 nm of the edges of $MoS_2$ on insulating substrate is discovered. This technique will be of great use for both fundamental and applied studies of various electronic devices.

## I. INTRODUCTION

Band diagrams are ubiquitously used to describe electronic devices. [1] However, most of the band diagrams have been drawn schematically or calculated theoretically based on the understanding of individual bulk materials that constitute the devices. With shrinking device sizes, band structure at the nanometer scale can have huge influence on the property of a device. As a result, direct measurement of the band profile with nanometer resolution is of great importance.

Scanning tunneling spectroscopy (STS) based on scanning tunneling microscope (STM) is the main technique to measure band profiles at the atomic scale. The working principle [2] of an STM is shown in Figure 1A. A metallic tip is brought close to the surface of a sample with a vacuum gap of ~ 1 nm. A tunneling current $I_t$ is measured at a bias voltage $V_b$. This current is used as the feedback signal to stabilize the vacuum gap. In the STS mode, the feedback is switched off and $V_b$ is swept. $I_t$ is recorded as a function of $V_b$, which contains information of local density of states (LDOS), hence the band structure under the tip can be measured. Since the STM needs tunneling current as the feedback signal and STS needs a stable vacuum gap after the feedback is off, they are usually limited to the interior of large (mm size) conductive samples at cryogenic temperatures. Modern electronic devices with micrometer or smaller sizes operated at room temperature pose great challenges for conventional STS. At room temperature, the tunneling current becomes noisy when the feedback is turned off, as shown in Figure S1 of the Supplementary Material (SM). On the other hand, devices based on novel quantum systems (such as quantum Hall effect [3] and quantum spin Hall materials [4]) have nontrivial edge states dominating their transport. Spatially-resolved band profiles at edges between a conductive material and an insulating substrate is important for the understanding of the intriguing physics. However, it is impossible to use conventional STS due to the absence of tunneling current on the insulating substrate.

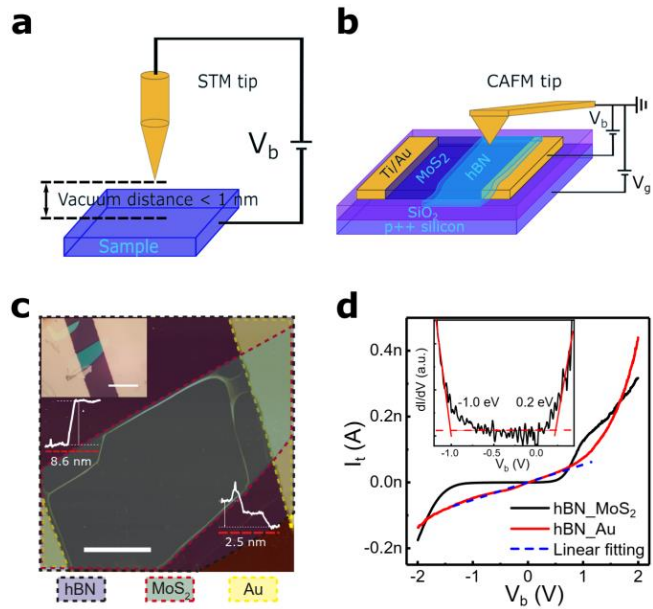

**Figure 1.** Comparison of conventional STS and CMSTS. (a) Schematic of the conventional STM. (b) Schematic of CMSTS. Both $V_g$ and $V_b$ are referenced to the tip, which is grounded. $V_g$ is applied to the gate and $V_b$ is applied to the sample. (c) The AFM image of a $MoS_2$ FET device covered by hBN. The white line runs along the edge of the device is an hBN wrinkle. Scale bar: 5 μm. The thicknesses of $MoS_2$ and hBN flakes are 8.6 nm and 2.5 nm, respectively. The inset of (c) is the optical microscope image corresponding to the main panel. Scale bar: 15 μm. (d) $I_t$ versus $V_b$ with tip on hBN/Au (red line) and hBN/$MoS_2$ (black line). The dashed line is linear fit to the low bias part of the red line. Inset: dI/dV to determine the band gap.

Here we develop a contact-mode STS (CMSTS) based on conductive atomic force microscope to remove the constraints of conventional STS. In our technique the tunneling junction is a wide band gap insulator instead of the vacuum gap, and the feedback signal is tip-sample contact force instead of tunneling current. The schematic of CMSTS is shown in Figure 1B, where a $MoS_2$ field-effect transistor (FET) is used as the model system to be measured. We demonstrate that CMSTS can be used to obtain band profiles of the $MoS_2$ device with nanometer resolution at room temperature. With the unique capability of imaging on insulating surfaces, we also find that the edges of $MoS_2$ flakes on the $SiO_2$ substrate have 0.6 eV band upshift with ~ 18 nm spatial extension.

## II. PRINCIPLE OF CMSTS

In CMSTS, a thin layer of hexagonal boron nitride (hBN) is inserted between the tip of a conductive atomic force microscope (CAFM) and the device as the tunnel junction (Figure 1B). Previous studies have established hBN as a superior insulator with wide band gap.[5] The layered structure makes it easy to get thin hBN flakes by exfoliation. One typical device is shown in Figure 1C, where a 2.5 nm hBN is transferred on top of a $MoS_2$ FET (see the Methods in SM for fabrication details). During measurement, the tip is stabilized by the contact force. All the experiments in this study were performed under ambient conditions as opposed to ultrahigh vacuum and low temperatures in most of the conventional STS.

We verify that CMSTS can measure the LDOS underneath the hBN in Figure 1D. The red curve is tunneling current $I_t$ versus sample bias $V_b$ on hBN/metallic electrode with tip grounded. At low bias within ± 1 V, $I_t$ varies linearly with $V_b$ as expected. Tunneling theory of the current between two metals with featureless LDOS separated by a constant tunnel barrier gives [6]

$$I_t(V_b) = \frac{A_{eff}\sqrt{m\varphi_B}q^2}{h^2 d} e^{-4\pi\sqrt{m\varphi_B}d/h} V_b, \quad (1)$$

where $A_{eff}$ is the effective tunneling area, $m$ is the electron mass, $\varphi_B$ is the tunnel barrier height of hBN, $q$ is the electron charge, $d$ is the hBN thickness and $h$ is the Plank constant. Previous study [7] has measured $\varphi_B$ of hBN to be 3.07 eV, which can be used to extract $A_{eff} \approx 0.25$ μm². Since the actual tip end is ~ 100 nm in diameter (see Figure S2), the large $A_{eff}$ could have two contributions. One is overestimate due to the uncertainty in the thickness of hBN, which will be discussed later. Second it may come from the possible water meniscus at the tip/hBN contact as observed before, [8] as the measurement is performed at ambient condition. Remarkably, the spatial resolution of band profiling is orders of magnitude higher than both the calculated $A_{eff}$ and the actual tip-end size as we will show later. At higher bias, the $I_t$ versus $V_b$ deviates from the linear dependence, which is due to the deformation of tunnel barrier and the emergence of Fowler-Nordheim tunneling process. [1] This nonlinearity also exists in conventional STS at bias outside of the ± 1 V range. [2] The asymmetry in Figure 1D can be attributed to the non-uniform electric field in the tunnel junction. [9]

When the $I_t$ versus $V_b$ is measured on hBN/MoS$_2$, the behavior dramatically changes (black curve in Figure 1D). There is negligible $I_t$ from ~ – 1 V to ~ + 0.5 V and strong increase outside this voltage range. When a numerical differentiation is performed (inset of Figure 1D), a curve closely resembles the LDOS measured by conventional STS on MoS$_2$ [10] is obtained. We assign the flat part in the d$I$/d$V$ curve to be the band gap, the upturn at the negative $V_b$ to be the valence band (VB) of MoS$_2$ and that at the positive side to be the conduction band (CB). Then the Fermi energy is about 0.2 eV from the CB edge and the band gap is 1.2 eV (see the Methods in SM for fitting details). These values agree very well with previously reported STS data on bulk MoS$_2$. [10]

In our measurements, the force applied between the tip and the sample is 300 nN. We have checked the dependence of the measured band structure on the contact forces (Figure S3). From 75 nN to 1500 nN, the locations of the CB and VB edges extracted from the CMSTS data has less than 0.1 eV variation, demonstrating the robustness of this method. Since the band gap of $MoS_2$ depends on its layer numbers, CMSTS is also used to measure the LDOS on monolayer $MoS_2$ (Figure S4), and the result closely agrees with that of conventional STS.[11] To emphasize on the crucial role played by the hBN layer in CMSTS, we also measured *I-V* directly on $MoS_2$ on the same device (Figure S5). Without a tunnel junction, the tip forms Ohmic or Schottky contact with the semiconductor and no LDOS information can be extracted.

Due to the ambient working condition of CMSTS, electric contacts to the device are easily accessible, which offers great flexibility to the experiment. We sweep the gate voltage $V_g$ between – 40 V and + 30 V, and track the evolution of $I_t$-$V_b$ curves (Figure 2A) in another device. When $V_g$ is swept to the positive side, the CV and VB currents show no obvious change. When $V_g$ is swept to the negative side the VB current remain the same, however, the CB current is strongly suppressed together with the band edge shifting towards higher $V_b$. This trend can be better seen in the contour plot of log ($dI_t$/$dV_b$) versus $V_b$ in Figure 2B. In the whole $V_g$ range, the VB shows little change, while the CB has obvious $V_g$ dependence in the negative side of $V_g$, leading to an apparent increase of the band gap. This behavior has been observed on multiple devices (see Figure S6 for the data from another device), and can be understood based on the band diagrams in Figure 2C and 2D. At zero $V_g$, tunneling current is measured when the tip Fermi level is aligned with the CB and VB edges (Figure 2C). Increasing $V_g$ to the positive side, the n type $MoS_2$ channel is in the accumulating mode, which can be inferred from the corresponding transport data in Figure 2E. The calculated Debye length is ~ 2 nm (see the Methods in SM for details), which means that the top layer of this $MoS_2$ flake (7 nm thick) is screened from the electric field produced by $V_g$. Since tunneling spectroscopy is surface sensitive, no change in the CMSTS data is expected. On the other hand, negative $V_g$ turns the $MoS_2$ channel to depletion mode. The depletion width can be much longer than the Debye length so the top layer of $MoS_2$ also feels the $V_g$ (see the Methods in SM). As shown in Figure 2D, the CB is pushed up by the negative $V_g$ which is observed in Figure 2B. However, although the VB is also pushed up in the channel, due to Fermi level pinning at the electrode/$MoS_2$ interface, [12] the hole current is blocked near the VB edge. The Fermi level of the tip needs to be lowered all the way to below the VB edge at the interface, which is not movable by $V_g$. We note that the CMSTS data in Figure 2A and 2B can also be interpreted as transfer curves of an FET with one electrode replaced by the tip and tunneling junction. In conventional FET transfer measurement of the same device (Figure 2E), only the electron conduction can be accessed which is generally explained by the Fermi level pinning and large Schottky barrier to the VB. [12] Here with the CMSTS technique, we can tunnel into the VB and directly show the existence of such a barrier at the contacts.

We should pay attention in this setup that the tunneling current is determined not only by the tunneling junction, but also by the lateral transport in $MoS_2$, especially at the metal/ $MoS_2$ contact. When the device is deeply turned off by the gate voltage, the contact resistance becomes comparable to or even larger than the tunneling resistance, then the voltage applied between the tip and the electrode is divided by the tunnel junction and the contact. The raw I-V data in that case will not reflect the real band information without careful calculation of the actual bias voltage taken by the tunneling junction. So when measuring tunneling spectroscopy of semiconductor devices with CMSTS, the gate voltage should be used to tune the transparency of the contacts in a reasonable range to avoid the complication of this issue.

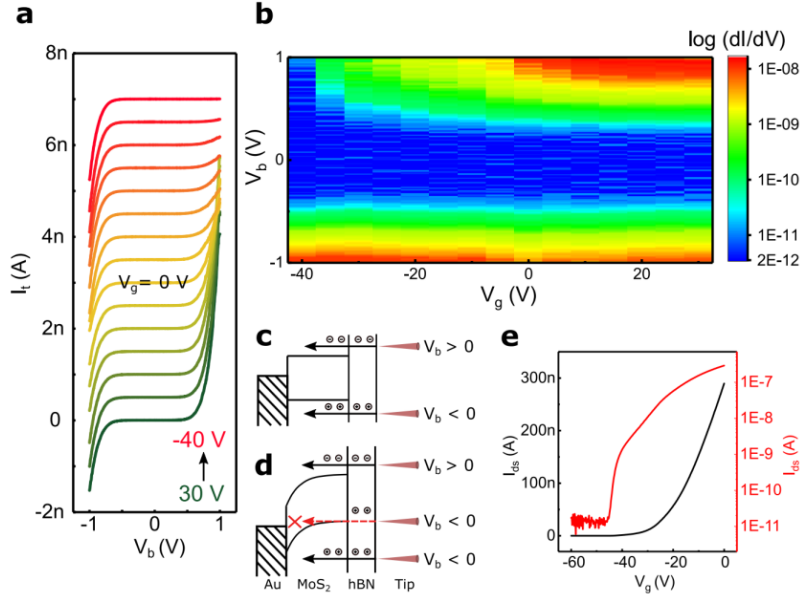

**Figure 2.** Tunneling spectroscopy tuned by the gate voltage. (a) CMSTS curves on hBN/MoS$_2$ as the back gate voltage is tuned from 30 V to – 40 V with a step of 5 V. (b) Contour plot of log (d$I$/d$V$) as a function of sample bias and gate voltage. (c) Band diagram at zero gate voltage. (d) Band diagram at a negative gate voltage, showing that the holes tunneling into the VB edges are blocked at the contact. (e) Transfer curve of the FET device in (a), displayed in linear (black) and log (red) scales. $I_{ds}$ is the drain-source current through the device.

We have also tried to directly measure the band profile at the metal/MoS$_2$ interfacing area. However, due to the very rough edge of the metal electrodes at the nanometer scale, no good profiling could be obtained.

## III. IMAGING THE EDGE STATES ON INSULATING SUBSTRATE

After establishing the validity of the CMSTS, we explore its unique capability of imaging at the conductor-insulator interface. Figure 3A shows an AFM topography image of the edge of a MoS$_2$ FET device (covered with hBN). A series of *I-V* spectroscopy are taken at every 6 nm along a line across the edge (red dashed line in Figure 3A), which are shown in Figure 3B. Compared to the interior of the MoS$_2$ flake, the edge has an obvious band up bending of about 0.6 eV, presumably due to the dangling bonds induced electron trapping centers.[13] The surface trapping density at the edge can be calculated to be $1.8 \times 10^{13}$ cm$^{-2}$eV$^{-1}$ (see the Methods in SM for details), which is typical for a semiconductor with a surface cutting through its covalent bonds.[1] We quantitatively extract the energies of CB minima and VB maxima (the red and black dots in Figure 3B), which can be fitted nicely with the model of a one-side abrupt junction (the red and black solid curves).[1] The edge is regarded as a reservoir of large number of electronic states. The depletion width and hence the spatial extension of the band bending is ~ 18 nm.

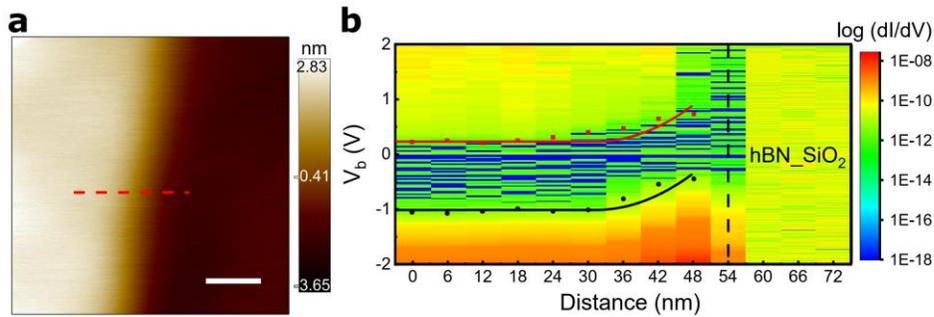

**Figure 3.** Detecting the band profile at the edge. (a) The AFM image of the MoS$_2$ edge (covered by hBN). Scale bar: 40 nm. Spectroscopies are taken along the red dashed line every 6 nm. (b) The contour plot of log (d*I*/d*V*) along the red dashed line in (a). The red and black dots are experimentally extracted CB and VB edges, respectively. The red and black solid curves are calculated band edges using the one-side abrupt junction model. The black dotted line shows the edge position.

Previous conventional STS measurements on the edges of $MoS_2$ [11] also revealed a band up bending of ~ 0.6 eV, but those experiments can only be performed on CVD grown $MoS_2$ flakes on graphite or other conducting substrates. Most of the high quality electronic devices based on $MoS_2$ use mechanically exfoliated flakes on insulating substrates, to which our results are of great relevance. For a real device with small size [14] or quantum transport, [15] the edges can even play a dominant role in the device property, and the understanding of band profile at the interface between the conductive channel and the insulating substrate is critical. For these systems the CMSTS has the unique capability compared with the conventional STM and STS.

**IV. SPATIAL RESOLUTION OF CMSTS**

Next we address the spatial resolution of the CMSTS. In Figure 4A, the 3D plot of an AFM topography at the $MoS_2$ edge is shown. Due to the size of the tip end (Figure S2), the atomically sharp edge appears to have a spatial extension of ~ 100 nm. At the same time of topography scan, the tunneling current is mapped. Remarkable images are obtained at – 2 V (Figure 4B) and + 2 V (Figure 4C) sample biases. Two intriguing features are prominent. First, when switching the polarity of sample bias, the edge current is changed from being enhanced to being suppressed, which are better seen in the line cuts (Figure 4D) taken from the 3D plots. This interesting phenomenon is a direct result of the band profile measured in Figure 3B. Due to the band up bending at the edge, tunneling to the VB is facilitated while to the CB is suppressed. Second and more importantly, the current mapping has a much finer spatial structure compared with that in the topography image of Figure 4A. In the line cut taken from the 3D plot in Figure 4B (middle curve of Figure 4D), the spatial extension of the edge is less than 10 nm, much narrower than its topography scan and much better than the calculated effective tunneling area $A_{eff}$ (0.25 μm$^2$).

Since $A_{eff}$ is exponentially dependent on the barrier height of hBN and the distance between the tip and MoS$_2$, which are not know accurately, $A_{eff}$ might be overestimated. Eq. (1) can be used to get $I_t(V_b) \propto \sim e^{-1.8d}$, where $d$ is measured in angstroms. If the $d$ is varied by 3.8 angstroms the current would change by 1000 times, which could account for the difference between the calculated $A_{eff}$ and the spatial resolution. Since the thickness of hBN was measured on SiO$_2$, it could be different from that on MoS$_2$.

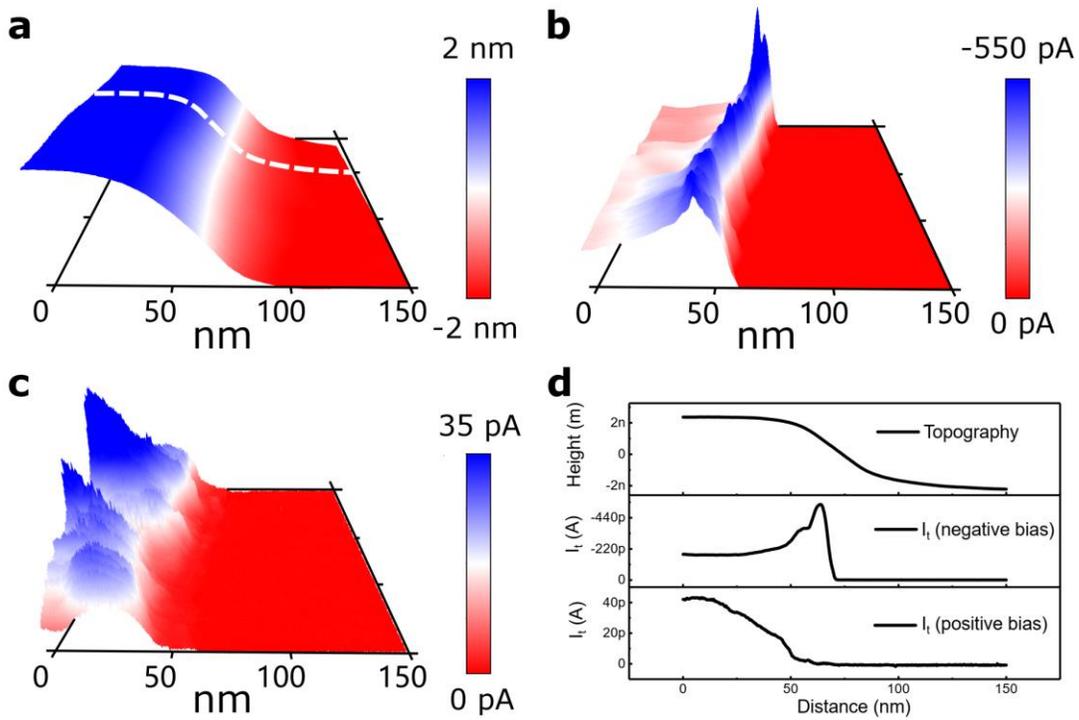

**Figure 4.** Spatial resolution of CMSTS. (a) The 3D plot of the AFM topography image at the MoS$_2$ edge (covered with hBN). The white dashed line is where the line profiles in (d) are taken. (b) and (c) 3D plots of the tunneling current images at sample bias of − 2 V and + 2 V, respectively. (d) The comparison between the height and current profiles along the white dashed line in (a).

We note that many intriguing edge states, such as the quantum Hall or quantum spin Hall edge states, has a spatial width of a few nanometers, [3, 4, 16] located at the interface between a nontrivial channel material and the trivial insulator (the substrate or vacuum). Some important semiconductors, such as two-dimensional hybrid perovskites, have edges with band gaps different from that of the interior and may dominate the carrier transport. [17] Previously, many techniques were used to map the edge states directly on insulating substrates, such as the scanning SQUID, [18] scanning microwave impedance microscopy, [19] scanning single electron microscopy, [20] photoluminescence mapping [17] and so on. Compared with CMSTS, these techniques have some limitations. First, their spatial resolution is usually in the range of micrometers, which is not high enough to detect some fine structures of the sample. Second, they usually cannot directly provide density-of-states information. For example, the scanning SQUID detects the magnetic field distribution induced by local current densities. [18] Scanning microwave impedance microscopy detects the microwave reflected by free carriers under the tip, so it measures the total carrier density instead of density of states. [19] Scanning single electron microscope detects the electric field distribution in space [20] and photoluminescence mapping [17] detects the local optical band gap. For CMSTS, we can obtain the information of local density of states, including both the occupied states and unoccupied states. We can also determine the Fermi level and band gap, and then the doping type and carrier density can be inferred. Together with the high spatial resolution and capability of scanning on insulating substrates, CMSTS can provide valuable information of the electronic properties of various materials and devices. The CMSTS technique developed here with nanometer resolution opens a door for exploring these exciting systems.

The data in Figure 4 also reveals another advantage of CMSTS compared with conventional STM: the topography and LDOS information are separated as shown in Figure 4A and 4B. In conventional STM, the tip-sample separation is determined by the tunneling current, which is influenced by both the physical height and LDOS of the sample. So the topography and LDOS information are convolved in an STM scan, no matter if it is nominally a topography scan or a STS mapping. In the CMSTS, however, topography and LDOS information do not interfere. Figure 4B shows a large tunneling current (i.e. large LDOS) at the edge of $MoS_2$, while the topography in Figure 4A shows no height increase there. This way, the topography reflects the physical height and CMSTS signal can be quantitatively compared throughout the sample since the tunneling distance is always the same.

Recently developed non-contact AFM/STM hybrid system [21] can also be used to disentangle the topography and LDOS. However, that is a much more specialized tool as compared with the common AFM used in CMSTS, and usually requires cryogenic temperature to operate. The vacuum tunneling gap between the tip and the sample in these hybrid systems shares the same stability problems of a conventional STM.

**V. CONCLUSION**

In conclusion, we have developed the new method (CMSTS) which can be used to obtain the band profiles of electronic devices with nanometer resolution. By changing the feedback mechanism from tunneling current in the conventional STS to contact force, CMSTS brings three major advantages. The first and most important one is the capability of measuring small (micrometer or smaller) devices and especially their edge states on insulating substrates. The second is the disentanglement of topography and LDOS information. In conventional STM the apparently high part of the scan does not necessarily correspond to topography height, but could originate from high LDOS; while in CMSTS, height scan has little influence from LDOS. The third is the stability of the tunneling gap, which could reduce noise in the tunneling current.

The most significant trade-off with the CMSTS compared to the conventional STM is the reduced spatial resolution. Since multiple atoms at the tip apex are tunneling, atomic resolution is not possible. However, for many devices and edge states, nanometer resolution could be enough. The increased tip size as compared with conventional STM also contributes to the stability of tunneling current. In conventional STM, the slight movement of the atoms at the tip apex will have huge impact on the tunneling current. Here, an area of nanometer size at the tip apex participates in the tunneling process, which makes the tunneling current more stable.

Finally, the hBN thickness uniformity, stability and conformity between hBN and the sample are critical, since the tunneling current sensitively depends on the distance between the tip and sample. To show that the tunneling current is stable and repeatable, we performed CMSTS measurements multiple times at a single spot (Figure S7A). Excellent stability and repeatability can be observed. We also measured tunneling current of the interior of a $MoS_2$ device covered by hBN (Figure S7B). A 20 nm × 20 nm area shows tunneling current variation of 23.77 pA on an average of 822.59 pA. This highly uniform current directly indicates the good conformity between hBN and the sample.

**Data Availability**

The data that support the findings of this study are available from the corresponding author upon reasonable request.

## SUPPLEMENTARY MATERIAL

Supplementary Material includes the device fabrication method, band gap fitting procedure, calculation of screening lengths, calculation of band profiles at the edge, calculation of surface trap density, and detail information of the AFM used in this work. It also contains supplementary figures 1 to 7.

## ACKNOWLEDGMENT

R.L. and J.X. acknowledge financial support from the Ministry of Science and Technology of China (No. 2016YFA0204000), the Strategic Priority Research Program of Chinese Academy of Sciences (No. XDA18010000) and ShanghaiTech University. K.W. and T.T. acknowledge support from the Elemental Strategy Initiative conducted by the MEXT, Japan, A3 Foresight by JSPS and the CREST (JPMJCR15F3), JST.

The authors thank the Soft Nano Fabrication Center at ShanghaiTech University.